\documentclass[aps,prb,twocolumn,showpacs,superscriptaddress,floatfix]{revtex4}

\usepackage{amsmath,amssymb,bm,graphicx}
\usepackage{xcolor}
\usepackage{siunitx}
\usepackage[colorlinks=true,urlcolor=blue,linkcolor=blue,citecolor=blue,bookmarks=false]{hyperref}

\begin{document}

\title{Dipolar effects on the critical fluctuations in Fe: Investigation by MIEZE}

\author{J. Kindervater}
 \affiliation{Physik-Department, Technische Universit\"at M\"unchen, D-85748 Garching, Germany}
 \affiliation{Institute for Quantum Matter and Department of Physics and Astronomy, Johns Hopkins University, 3400 North Charles Street Baltimore, MD 21218, USA}
 
\author{S. S\"aubert}
 \affiliation{Physik-Department, Technische Universit\"at M\"unchen, D-85748 Garching, Germany}
 \affiliation{Heinz Maier-Leibnitz Zentrum, Technische Universit\"at M\"unchen, Lichtenbergstr. 1, D-85748 Garching, Germany}

\author{P. B\"oni}
 \affiliation{Physik-Department, Technische Universit\"at M\"unchen, D-85748 Garching, Germany}

\date{\today}

\begin{abstract}
Iron is one of the archetypical ferromagnets to study the critical fluctuations at a continuous phase transition thus serving as a model system for the application of scaling theory. We report a comprehensive study of the critical dynamics at the transition from the ferro- to the paramagnetic phase in Fe, employing the high-resolution neutron spin echo technique MIEZE. The results show that the dipolar interactions lead to an additional damping of the critical spin fluctuations at small momentum transfers $\bf q$. The results agree essentially with scaling theory if the dipolar interactions are taken into account by means of the mode-coupling equations. However, in contrast to expectations, the dipolar wavenumber $q_D$ that plays a central role in the scaling function $f(\kappa/q,q_D/\kappa)$ becomes temperature dependent. In the limit of small $\bf q$ the critical exponent $z$ crosses over from 2.5 to 2.0. 
\end{abstract}
\vskip2pc
\maketitle

\section{Introduction}
\label{sc:ironintroduction}

The concept of scaling is an important tool for the classification of phase transitions. In their seminal work on dynamic critical phenomena, Hohenberg and Halperin \cite{1977:hohenberg:RevMosPhys} demonstrated many decades ago that the dynamic critical exponents describing the relaxation of the flucutuations near a second order phase transition can be classified according to the spatial dimensionality, the component number of the order parameter, the range of the interactions, and the conservation of the order parameter.

Magnetic phase transitions are ideally suited to study crossover phenomena between different universality classes near critical points, in contrast to structural phase transitions involving soft modes, where the order parameter is usually strongly coupled to the lattice leading to a first order transition. In most magnetic systems the coupling between the order parameter and the fluctuations can be neglected. Therefore the magnetic phase transitions remain second order allowing to measure the critical fluctuations and to determine the critical exponents even very close to the phase transition. A notable exception of this scenario is the fluctuation induced first order phase transition from the helimagnetic to the paramagnetic phase in MnSi that can be explained in terms of a Brasovskii phase transition \cite{2013:Janoschek:PhysRevB,2014:Kindervater:PhysRevB}. Important tools to interpret the critical dynamics at continuous phase transitions are beside dynamical scaling theory \cite{1967:Halperin:PhysRevLett,1969:Halperin:PhysRev}, mode coupling theory \cite{1960:Fixman:JChemPhys,1962:Fixman:JChemPhys} and renormalization group theory \cite{1973:Fisher:PhysRevLett}.

Dynamical scaling theory predicts \cite{1977:hohenberg:RevMosPhys} that the inverse lifetime of the critical fluctuations in an isotropic ferromagnet at the Curie temperature $T_C$ is given by
\begin{equation}
	\Gamma = f(\kappa/q)Aq^z,
\label{dynscal}
\end{equation}
where $f(\kappa/q)$ is a dynamical scaling function, $\kappa = 2\pi/\xi$ the inverse correlation length, $z = 2.5$ the critical dynamic exponent, and $A$ a material-specific constant that expresses the energy scale of the exchange interactions. We have neglected the Fisher exponent $\eta = 0.034$ \cite{1977:LeGuillou:PhysRevLett}. Although $z = 2.5$ was essentially confirmed by neutron scattering in the itinerant magnets Fe \cite{1969:Collins:PhysRev} and Ni\cite{1969:Minkiewicz:PhysRev} and in EuO \cite{1986:Boeni:PhysRevB}, deviations from the expected behavior became apparent in the model Heisenberg ferromagnets EuO and EuS\cite{1976:Passell:PhysRevB}. Later, the non-universality of Eq. (\ref{dynscal}) was also demonstrated in Fe using the neutron spin echo technique where marked disagreements were observed at small momentum transfers $\bf q$ \cite{1982:Mezei:PhysRevLett,1984:Mezei:JMagnMagnMater,1986:Mezei:PhysicaB}. Deviations have also been observed in Ni \cite{1993:Boeni:PhysRevB}.

It was Frey and Schwabl \cite{1987:Frey:PhysLettA,1988:Frey:PhysLettA,1988:Frey:ZfPhysik} who clarified the characteristic deviations of the measured dynamic scaling function from the isotropic behavior. They evaluated numerically the mode-coupling (MC) equations for a Heisenberg ferromagnet including the dipolar interactions and were able to predict the observed cross-over from isotropic critical behavior to dipolar critical behavior in EuS at small $q$ \cite{1991:Boeni:PhysRevB} by inclusion of the dipolar wavenumber $q_D$ in the dynamical scaling function, i.e. 
\begin{equation}
	\Gamma = f(\kappa/q, q_D/\kappa)Aq^z.
\label{dipscal}
\end{equation}
Note that $q_D$ can be inferred from the relation between $\kappa(T)$ and the homogeneous internal susceptibility $\chi(T)$ using the expression $\chi = (q_D/\kappa)^2$ or from the saturation magnetization $M_S$ and the stiffness $D$ of the magnons by the relation $\mu_0\mu_BM_S = Dq_D^2$. The polarization of the paramagnetic fluctuations with respect to $\bf q$ is neglected in Eq. (\ref{dipscal}) as only transverse fluctuations $\delta {\bf S} \perp \bf q$ are observed in the experiment discussed below. 
On the one hand, the characteristic deviations of the paramagnetic fluctuations in Fe \cite{1982:Mezei:PhysRevLett} from Eq. (\ref{dynscal}) seemed to give further credit to the MC-results while on the other hand the deviations of the measured $f(\kappa/q)$ from the isotropic behavior in Ni\cite{1993:Boeni:PhysRevB} and Pd$_2$MnSn \cite{1986:Kohgi:PhysRevB,1989:Graf:PhysRevB} may not be compatible with the MC-theory.

In order to confirm the predictions of MC-theory in a ferromagnet, we have investigated the critical dynamics in Fe using the neutron spin-echo technique MIEZE\cite{1992:Gaehler:PhysicaB} in the longitudinal DC field configuration\cite{2016:Krautloher:RevSciInstr}. Fe is ideally suited for these experiments because it exhibits a large moment $\mu_{Fe} = 2.1 \mu_B$ and a reasonably small stiffness $D = 281\pm10$ meV\AA$^{2}$ \cite{1969:Collins:PhysRev} at 295 K leading to a dipolar wavenumber $q_D = 0.033$ \AA$^{-1}$\,\cite{1986:Kotzler:JMagnMagnMat} that is well in the range for small angle neutron scattering (SANS). In contrast, Ni is not suited as well because of its small moment $\mu_{Ni} = 0.6 \mu_B$ and the large stiffness $D = 555$ meV\AA$^{2}$ \cite{1973:mook:PhysRevLett} at 4.2 K. The results show that in contrast to the localized ferromagnets such as EuS a heuristic temperature dependence of the dipolar wavenumber $q_D(T)$ may be introduced reflecting the interaction of the spin fluctuations with the conduction electrons.

\section{Experimental results}
\begin{figure}
	\centering
	\includegraphics[width=0.48\textwidth]{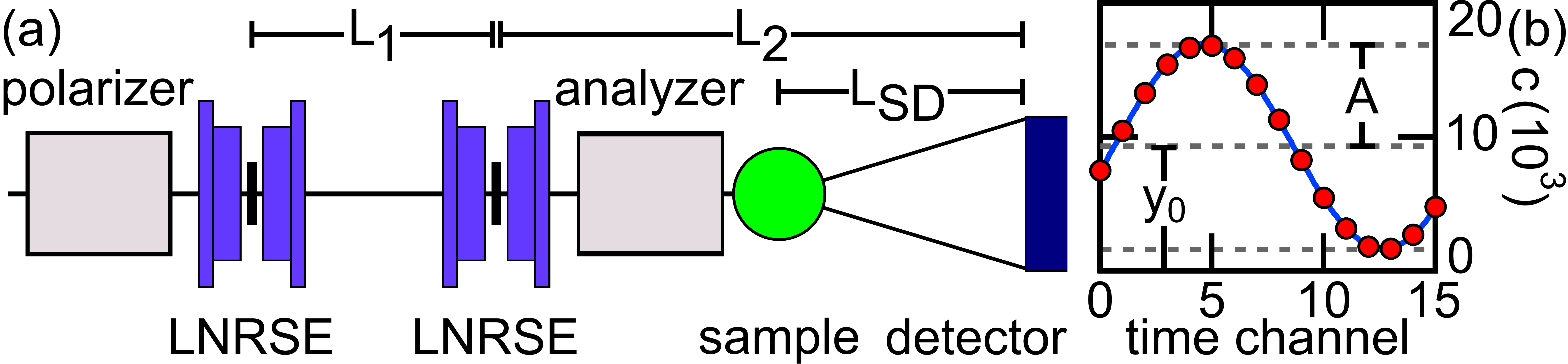}
	\caption{(a) Sketch of the MIEZE setup used in the experiment. 
	Neutrons are traveling the spectrometer from left to right. 
	The neutron spin is manipulated by a longitudinal (i.e. in beam direction) DC magnetic field and a radio frequency field perpendicular to the DC field in so called LNRSE coils.
	All spin manipulation takes place in front of the sample, such that spin manipulation a the sample position does not influence the performance of the spectrometer. 
	(b) Typical MIEZE echo (red circles) fitted by a cosine function (blue line) of the form $I=A\cos(k\cdot t + \phi_0)+y_0$, with the amplitude $A$, a parameter $k$ proportional to the MIEZE time $\tau$, an additional constant phase $\phi_0$, and the average counts $y_0$.}
	\label{fig:fig1}
\end{figure}
The high-resolution measurements were performed at the beamline RESEDA\cite{2007:Haeussler:PhysicaB} at the MLZ, using the newly developed longitudinal DC field MIEZE option\cite{2016:Krautloher:RevSciInstr,2015:Kindervater:EPJWebConf}. 
For our experiments we used neutron wavelengths $\lambda=\SI{5.4}{\angstrom}$ and $\lambda=\SI{8.0}{\angstrom}$. 
The distances between the longitudinal neutron resonance spin echo (NRSE) coils, between the second NRSE coil and the detector, and between sample and detector were $L_1=\SI{1.926}{\meter}$, $L_2=\SI{3.818}{\meter}$, and $L_\textrm{SD}=\SI{2.525}{\meter}$, respectively, as sketched in Fig~\ref{fig:fig1}(a). 
In this configuration a dynamic range from \SI{1.6e-5}{\nano\second} to \SI{5}{\nano\second} was covered. 

The experiments were conducted on a single crystal of bcc iron. It has a diameter of 9 mm and a length of 25 mm and was previously used by Wicksted et al.  \cite{1984:Wicksted:PhysRevB}.
The $\langle 110 \rangle$ axis is aligned approximately \SI{10}{\degree} off the cylinder axis.
It was mounted vertically in a high temperature furnace, using a resistive Nb double cylinder heating element, allowing to heat the sample above the Curie temperature $T_\textrm{C}=\SI{1043}{\kelvin}$ \cite{1995:Kittel:book}.
A Eurotherm temperature controller was used to measure and control the temperature with a stability of $\Delta T\approx \SI{0.05}{\kelvin}$.

\subsection{Integrated small angle scattering}
\begin{figure}
	\centering
	\includegraphics[width=0.45\textwidth]{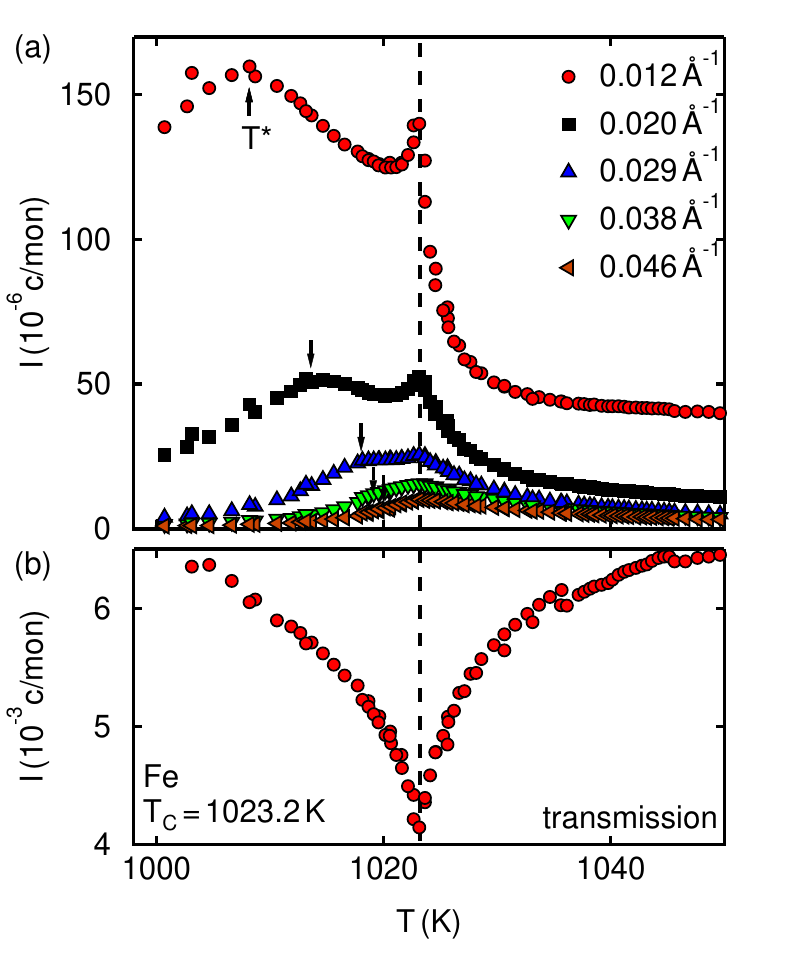}
	\caption{Temperature dependence of the critical magnetic scattering near the Curie temperature $T_\textrm{C}$.
	(a) Scattering intensity as function of temperature evaluated for different scattering vectors $q$.	
	The Curie temperature $T_\textrm{C}$ is marked by the dashed line, the peak at $T^*$ can be attributed to shrinking ferromagnetic domains.
	(b) Intensity of transmitted neutrons through the sample as function of temperature. 
	The Curie temperature $T_\textrm{C}$ is defined by the sharp minimum of the transmission and the maximum of the critical scattering.}
 	\label{fig:IronIntVsT}
\end{figure}

The Curie temperature $T_C$ of the sample was determined by measuring the temperature dependence of the critical magnetic scattering using RESEDA in the small angle neutron scattering (SANS) configuration. Fig. \ref{fig:IronIntVsT} shows the temperature dependence of (a) the scattered intensity evaluated for various momentum transfers $q$ and (b) the transmission through the sample. A sharp peak in the scattered intensity and a sharp minimum in the transmission at $T_\textrm{C}=\SI{1023.2}{\kelvin}$ define $T_C$ within \SI{\pm 0.1}{\kelvin}. The offset of about \SI{20}{\kelvin} to the value $T_C = 1043$ K reported in the literature of \SI{1043}{\kelvin} is due to a temperature gradient between the heater and the sample.

A temperature gradient inside the sample can be excludes as the peak width is very narrow, i.e. $\Delta T =\SI{0.2}{\kelvin}$. 
In addition, temperature scans with increasing and decreasing temperature do not show any hysteresis. 
The distinct suppression of the neutron transmission at $T_\textrm{C}$ is a vivid example for the phenomenon of critical opalescence.

The scattering intensity has a second, broad maximum at $T^*<T_\textrm{C}$ in the ferromagnetic phase. The peak shifts with increasing $T^*<T_\textrm{C}$ to larger $q$. This effect is attributed to the shrinking ferromagnetic domains when approaching $T_C$. A similar effect has been observed in EuO \cite{1976:AlsNielsen:PhysRevB}.

\subsection{Quasielastic measurements}
In order to extract the linewidth of the critical fluctuations near the phase transition, we have performed quasielastic measurements in a SANS geometry with a position sensitive CASCADE detector \cite{2011:Haussler:RevSciInstrum} centered at $Q=0$. 
In this configuration, only spin fluctuations with a transverse polarization with respect to the momentum transfer $\bf q$ are measured while longitudinal fluctuations do not contribute to the scattering cross section\cite{1991:Boeni:PhysRevB}.
The measurements were performed at temperatures $T_C \le T \le T_\textrm{C} + \SI{30}{\kelvin}$ covering a $q$-range of $\SI{0.013}{\per\angstrom} \leq q \leq \SI{0.068}{\per\angstrom}$ at $\lambda=\SI{5.4}{\angstrom}$ and $\SI{0.009}{\per\angstrom} \leq q \leq \SI{0.043}{\per\angstrom}$ at $\lambda=\SI{8.0}{\angstrom}$ with one experimental setting, respectively.\\

As in classical spin echo the polarization or contrast $C$ of a MIEZE echo is directly proportional to the intermediate scattering function $S(q,\tau)$.
The echo can be fitted as function of time $t_D$ by a simple cosine function where the intensity is given by $I=A\cos(k\cdot t_D)+y_0$, with the amplitude $A$, $k$ a parameter proportional to the spin echo time $\tau$ and the average intensity $y_0$.
Fig~\ref{fig:fig1}(b) shows typical data for a MIEZE echo (red circles) and a fitted curve (blue line). 
The intermediate scattering function can be recovered from the fit parameters by calculating $S(q,\tau)/S(q,0)=A(\tau)/y_0(\tau)$ for each recorded spin echo time $\tau$.
To account for the instrumental resolution and sample geometry we have divided all data by a resolution measurement using an elastically scattering graphite sample.
The resolution measurement was performed in the very same configuration as the iron measurement.

\begin{figure}[htb]
	\centering
	\includegraphics[width=.5\textwidth]{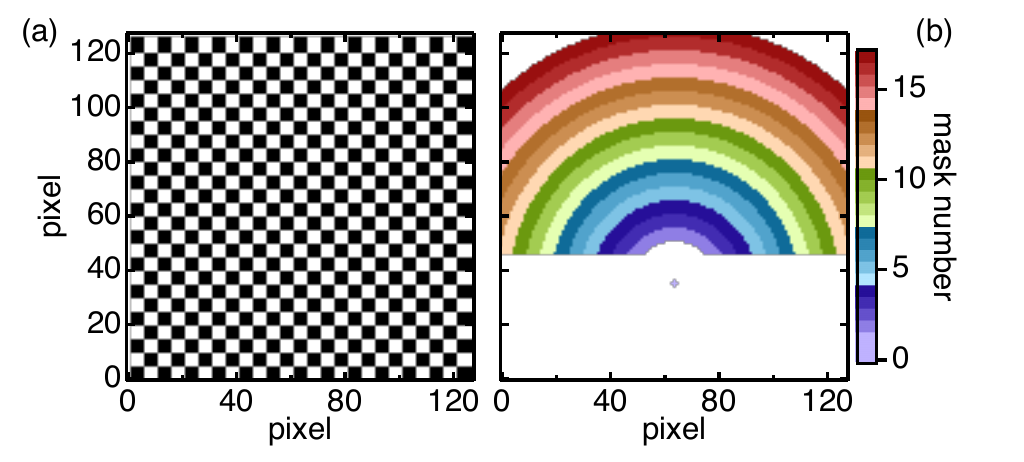}
	\caption{Pregrouping and postgrouping masks as used for the measurement of the magnetic fluctuations in Fe. In panel (a) the pregrouping mask is shown. The $128\cdot128$ pixels of the detector are combined in regions with a size of $5\cdot5$ pixels (black and white squares).
	The postgrouping, shown in panel (b) consists of circles centered at the direct beam. 
	Each slice has a width of 5 pixels.
	The pixels 0 - 45, corresponding to the large white region in (b) are not evaluated due to contamination of the data by background from the spin analyzer.}
	\label{fig:Iron_Masks}
\end{figure}

The data reduction process was as follows:
In a first step, the detector is \textit{pregrouped} using a mask summing the counts of arrays of $5\cdot5$\,pixels each as shown in Fig.~\ref{fig:Iron_Masks}(a).
The MIEZE echo is then fitted to these combined counts in order to extract the amplitude $A$ and average counts $y_0$, as defined in Fig~\ref{fig:fig1}(b), with sufficient statistics.
Finally, $A$ and $y_0$ are summed on rings of constant $Q$ or more precisely of constant scattering angle $2\theta$ as defined by the \textit{postgrouping} mask shown in Fig.~\ref{fig:Iron_Masks}(b).
The rings are centered around the direct beam with a width of 5\,pixels each. 
This procedure of \textit{pre}- and \textit{post}- grouping allows to extract the intermediate scattering function $S(q,\tau)$ with sufficient statistics, while still keeping a high MIEZE contrast.

\begin{figure*}
	\centering
 	\includegraphics[width=0.95\textwidth]{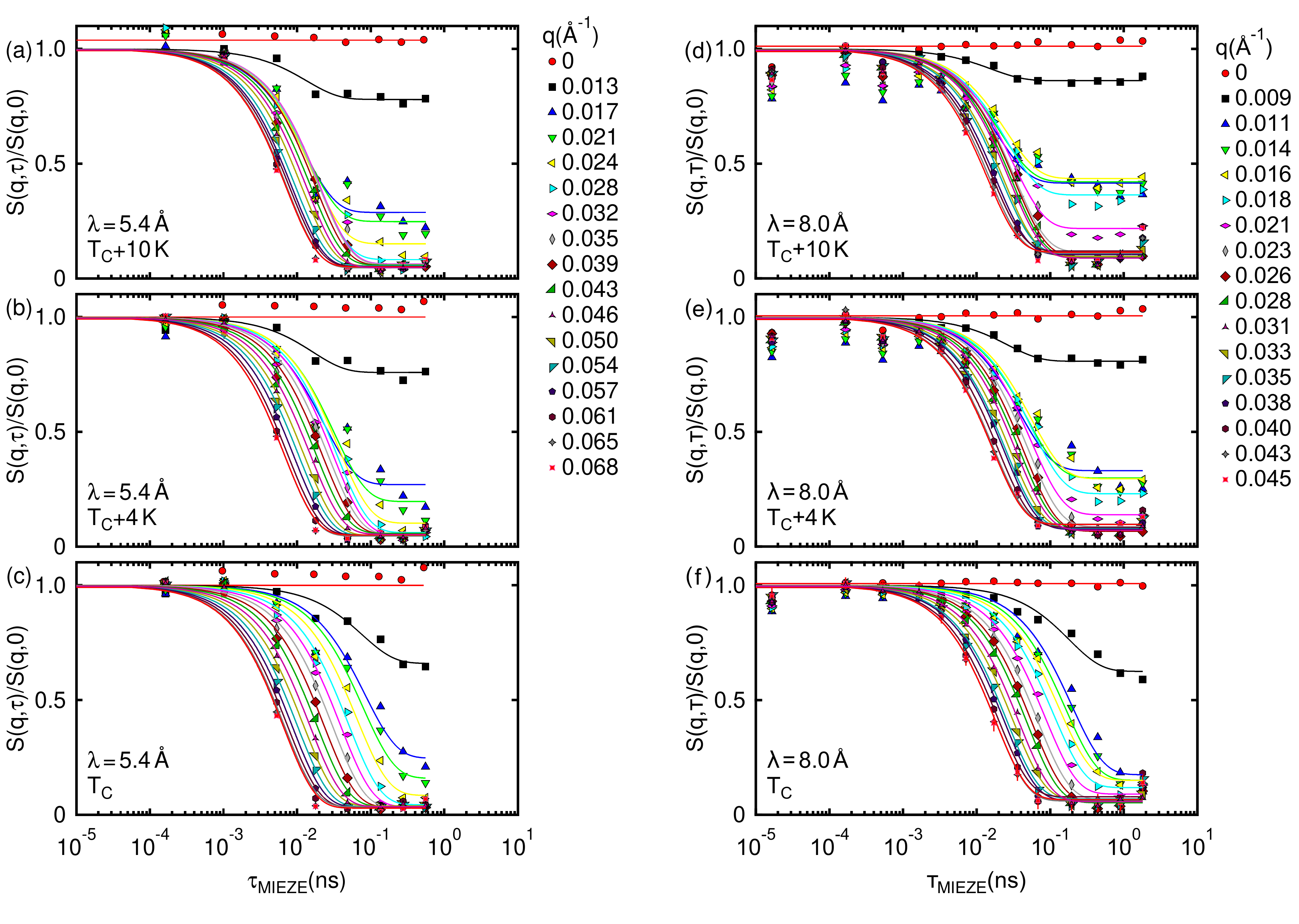}
 	\caption{Normalized MIEZE contrast in iron for measurements at (a), (d) $T_\textrm{C}+10$\,K, (b), (e) $T_\textrm{C}+4$\,K and (c), (f) $T_\textrm{C}$.
 	Data were recorded using neutrons with a mean wavelength of $\lambda=5.3$\,\AA (a)-(c) and $\lambda=8.0$\,\AA (d)-(f).
	The solid lines are fits to the data using  Eq.~(\ref{eq:GammaFitFreeA}).}
 	\label{fig:IronSvQTau_Lambda5p3}
\end{figure*}

Fig. \ref{fig:IronSvQTau_Lambda5p3} shows the normalized intermediate scattering function $S(q,\tau)/S(q,0)$ measured at the Curie temperature $T_\textrm{C}$, $T_\textrm{C}+\SI{4}{\kelvin}$ and $T_\textrm{C}+\SI{10}{\kelvin}$ using both a neutron wavelength of \SI{5.4}{\angstrom} and \SI{8.0}{\angstrom}, respectively. The life time of the fluctuations decreases with increasing scattering vector and increasing temperature, as expected due to critical slowing down, i.e. large patches of spins have a longer life time.

At small $q$, an offset in the intermediate scattering function is observed which is due to background. 
Moreover, in the direct beam at $q=0$ for $\lambda=\SI{5.4}{\angstrom}$ (cf. Fig. \ref{fig:IronSvQTau_Lambda5p3}) $S(q,\tau)/S(q,0)$ is slightly larger than one. 
This unphysical effect can be attributed to very large count rates in the direct beam. 
At $\lambda=\SI{8.0}{\angstrom}$, were the neutron flux is 12 times lower compared to $\lambda=\SI{5.4}{\angstrom}$, $S(q,\tau)/S(q,0)=1$ in the direct beam.\\

The spectrum of the critical fluctuations above $T_\textrm{C}$ is assumed to have a Lorentzian line shape \cite{1984:Shirane:JMagnMagnMater}, hence in time-space all data are fitted by assuming a single exponential decay of the form 
\begin{equation}
	\frac{S(q,\tau)}{S(q,0)}=(1-B)\cdot \exp\left[-\frac{\Gamma \tau}{\hbar}\right]+B,
	\label{eq:GammaFitFreeA}
\end{equation}
where $0\le B\le 1$ is a $q$-dependent parameter that allows to account for an elastic background due to the direct beam, $\Gamma$ the linewidth of the fluctuations, $\tau$ the spin echo time and $\hbar$  Planck's constant divided by $2\pi$.
The resulting fits reproduce the data shown in Fig. \ref{fig:IronSvQTau_Lambda5p3} well.\\

\begin{figure}
	\centering
	\includegraphics[width=0.45\textwidth]{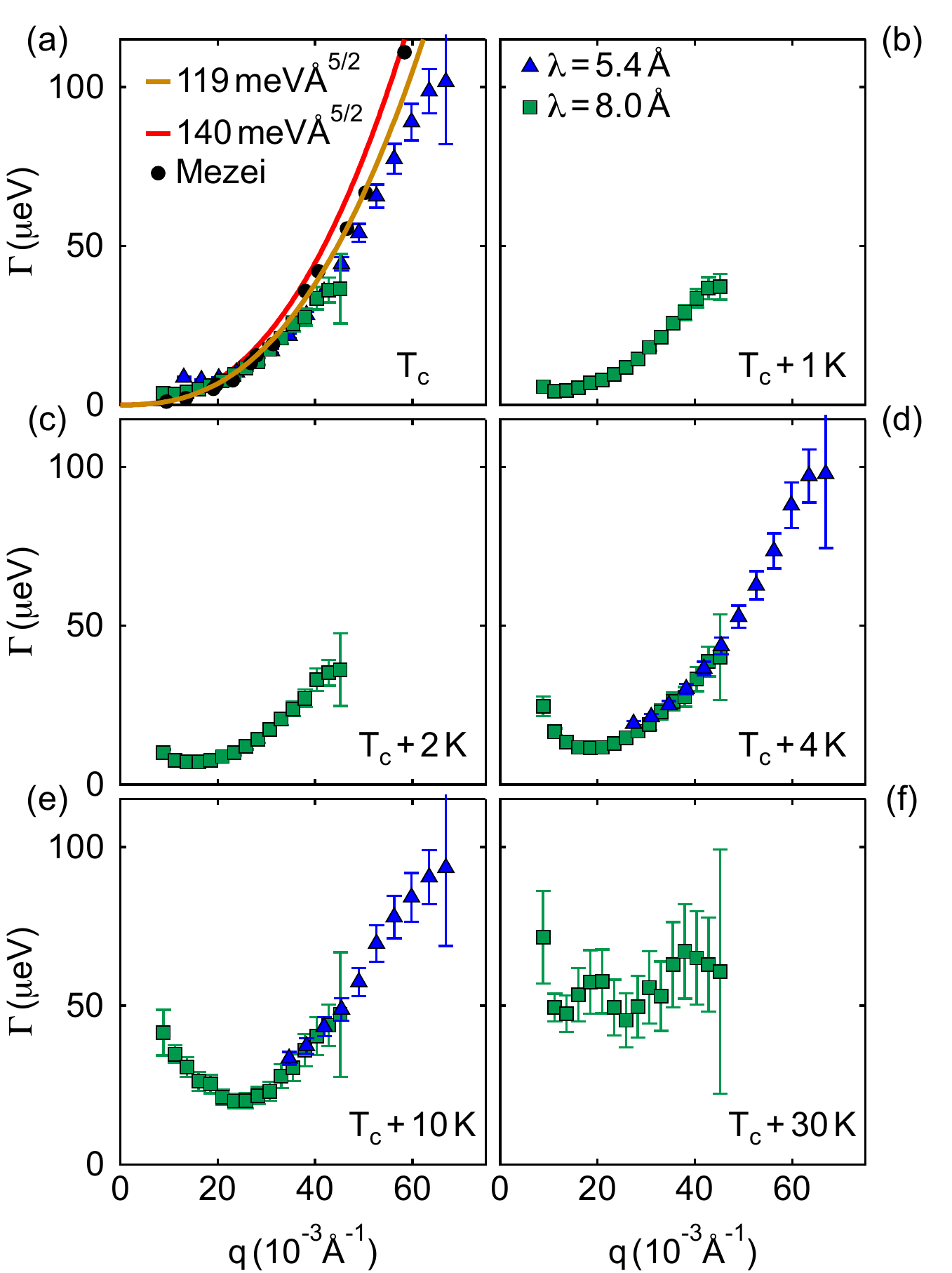}
	\caption{Linewidth of the critical fluctuations in iron as function of the scattering vector $q$, measured at different temperatures: $T=T_\textrm{C}$ (a), $T=T_\textrm{C}+\SI{1}{\kelvin}$ (b), $T=T_\textrm{C}+\SI{2}{\kelvin}$ (c), $T=T_\textrm{C}+\SI{4}{\kelvin}$ (d), $T=T_\textrm{C}+\SI{10}{\kelvin}$ (e), $T=T_\textrm{C}+\SI{30}{\kelvin}$ (f).
	The blue triangles are measured with a neutron wavelength of $\lambda=\SI{5.4}{\angstrom}$ and the green squares with $\lambda=\SI{8.0}{\angstrom}$, respectively.
	The black circles in panel (a) are the results reported by Mezei \textit{et al.} \cite{1982:Mezei:PhysRevLett}.
	Solid lines represents $Aq^{5/2}$ with $A=140$\,meV\AA$^{5/2}$ \cite{1984:Shirane:JMagnMagnMater} (red curve) and  $A=119$\,meV\AA$^{5/2}$ (orange curve).}
	\label{fig:IronGammaVsQ}
\end{figure}

The $q$-dependence of the linewidth $\Gamma$ for various temperatures is shown in Fig. \ref{fig:IronGammaVsQ}.
Blue triangles and green squares represent data as obtained with neutrons of wavelengths $\lambda = \SI{5.4}{\angstrom}$ and  \SI{8.0}{\angstrom}, respectively. 
For all temperatures an increase of $\Gamma$ with increasing $q$ is observed. The statistical errors increase with increasing $q$ due to significantly smaller count rates. 

In panel (a) the present data is directly compared with data as measured with the classical neutron spin echo (NSE) technique \cite{1982:Mezei:PhysRevLett} (black circles) and data as obtained by triple axis spectroscopy (TAS)\cite{1984:Shirane:JMagnMagnMater}. The results of all used techniques are in excellent agreement with each other. The solid lines are given by $\Gamma=Aq^{2.5}$, with $A=140$\,meV\AA$^{5/2}$ (red line)\cite{1984:Shirane:JMagnMagnMater}, and  $A=119\pm17$\,meV\AA$^{5/2}$ (fit to our data, orange line). These values of $A$ agree also well with $A = 140$ meV\AA$^{2.5}$ as obtained by triple axis spectroscopy (TAS)\cite{1984:Shirane:JMagnMagnMater}.
The difference between the TAS and the NSE data may be due to the following reason: The TAS-measurements have been conducted at fixed $q$ while in NSE and MIEZE, the measurements are performed at constant scattering angle.
Another origin of the difference could be due to the assumption of a Lorentzian line shape.
Very close to $T_\textrm{C}$ the line shape is no more Lorentzian as shown for EuO \cite{1987:Boeni:PhysRevB}.

We point out that all values $A$ are compatible with the theoretical value predicted by Frey and Schwabl \cite{1987:Frey:PhysLettA}, namely $A=128.6$\,meV\AA$^{5/2}$. 
Note that at small $q$ there is a distinct difference between the trend of the experimental results and the $q$-dependence of the linewidth at $T_C$, i.e. $\Gamma=Aq^{2.5}$ (Fig. \ref{fig:IronGammaVsQ} (b) - (f)).
The difference is a first indication for the influence of the dipolar interaction which will be discussed next.\\

\begin{figure}
	\centering
	\includegraphics[width=0.5\textwidth]{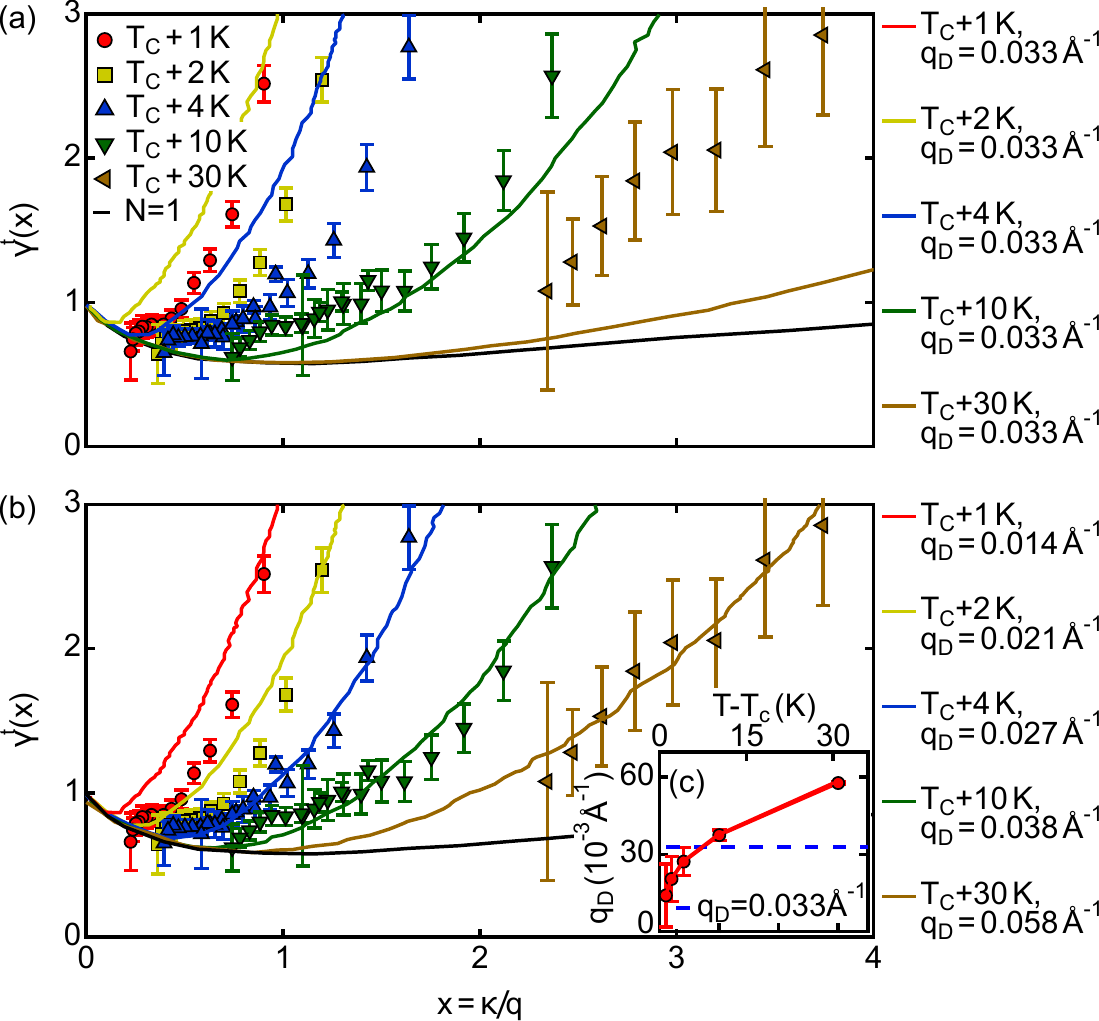}
	\caption{Scaling function of the critical fluctuations calculated from the linewidth shown in Fig. \ref{fig:IronGammaVsQ} using Eq.~\ref{eq:howtocalcscalenfunc} with $A=119$\,meV\AA$^{5/2}$.
	The solid lines are the scaling functions calculated within the theory by E. Frey and F. Schwabl \cite{1987:Frey:PhysLettA,1988:Frey:ZfPhysik,2008:Frey:arXiv}.
	Panel (a) shows the scaling function as calculated for a constant $q_D=\SI{0.033}{\per\angstrom}$. In panel (b) $q_D$ is adjusted to fit scaling function and experimental results.
	Panel (c) shows the temperature dependence of $q_D$ as used in panel (b).}
	\label{fig:SkalenFkt_Iron}
\end{figure}

The hypothesis of dynamical scaling, Eq.~(\ref{dipscal}), allows to extract the dynamical scaling function for different temperatures by calculating
\begin{equation}
 	f^T(x,q_D /\kappa)=\frac{\Gamma(T)}{\Gamma(T_\textrm{C})}=\frac{\Gamma(T)}{Aq^{2.5}},
 \label{eq:howtocalcscalenfunc}
\end{equation}
where $x=\kappa/q$. 
The superscript $T$ indicates that our experiment provides the dynamics for transverse spin fluctuations. 
Fig. \ref{fig:SkalenFkt_Iron}(a) shows a comparison of the scaling functions as obtained from our data using $A=119$\,meV\AA$^{5/2}$ (Fig.~\ref{fig:IronGammaVsQ} (b) - (f)) and from Frey and Schwabl
\cite{2008:Frey:arXiv} using a constant dipolar wavenumber $q_D=\SI{0.033}{\per\angstrom}$ as reported in Ref. \cite{1986:Kotzler:JMagnMagnMat}. The experimental and theoretical scaling functions $f^T$ exhibit a similar dependence on $x$, i.e. assuming a minimum in the range $0<x<1$ and increasing for larger $x$. Also the trend as function of temperature is similar. 

Nevertheless, the agreement between theory and experiment is not satisfactory for both low and high temperatures. In order to match the theoretical and experimental results for the scaling function we have allowed a variation of $q_D$ as function of temperature, as shown in Fig. \ref{fig:SkalenFkt_Iron}(b). Note that his procedure has to be considered as a heuristic parametrisation of the data. The thus obtained temperature dependence of the dipolar wave numbers $q_D$ is shown in (c), where the blue dashed line reflects the literature value $q_D=\SI{0.033}{\per\angstrom}$ from Ref. \cite{1986:Kotzler:JMagnMagnMat}. 
For $T_\textrm{C} \le T \le T_\textrm{C}+\SI{8}{\kelvin}$ $q_\textrm{D}$ is slightly smaller, while above $T_\textrm{C}+\SI{8}{\kelvin}$, $q_\textrm{D}$ is significantly larger than the literature value.

The systematic deviation of the dipolar wavenumber from the literature value $q_D = 0.033$ \AA$^{-1}$ may be understood as an additional damping of the transverse magnetic fluctuations by the conduction electrons.
Similar deviations of the damping have also been found in nickel as reported in Ref.~\cite{1993:Boeni:PhysRevB,1993:Boeni:PhysicaB} and in Pd$_2$MnSn\cite{1986:Kohgi:PhysRevB,1989:Graf:PhysRevB}.

\section{Conclusions}
\label{sc:iron_discussion}
Using the MIEZE technique we have investigated the influence of the dipolar interactions on the critical dynamics in the itinerant ferromagnet Fe above the Curie temperature $T_C$. 
The systematic deviations of the dynamic scaling function from the one expected for the isotropic critical behavior resemble very closely the deviations observed in the localized ferromagnet EuS\cite{1991:Boeni:PhysRevB}, which can be reproduced quantitatively by mode-coupling theory \cite{1987:Frey:PhysLettA,1988:Frey:PhysLettA,1988:Frey:ZfPhysik}. 
We have identified an additional damping mechanism in itinerant Fe that may be attributed to spin-flip excitations of the conduction electrons, which may originate from the large $T_C$ relativ to the Fermi temperature. Therefore, the Fermi surface is smeared out and will facilitate spin fluctuations at small momentum transfer $q$.

The experiments on Fe establish MIEZE to be a valuable and efficient technique for the investigation of critical dynamics in ferromagnets. 
In contrast to conventional spin echo techniques, MIEZE also allows measurements on depolarizing samples and samples exposed to magnetic fields\cite{2015:Kindervater:EPJWebConf,2011:Georgii:ApplPhysLett}.

\acknowledgements
We thank C. Pfleiderer for useful discussions and M. Antic for support with the high temperature furnace.
Financial support through DFG TRR80 (From Electronic Correlations to Functionality) and ERC AdG (291079, TOPFIT) is gratefully acknowledged. 
J.K., and S.S.\ acknowledge financial support through the TUM graduate school.
\bibliography{bib_critical_scattering_iron}
\end{document}